\documentstyle[aps,prbbib,twocolumn,epsf]{revtex}

\begin{document}
\draft
\title{Shot noise of spin current}

\author{Baigeng Wang$^{1,2}$, Jian Wang$^{1,3,a}$, and Hong Guo$^4$}
\address{1. Department of Physics, The University of Hong Kong, 
Pokfulam Road, Hong Kong, China\\
2. National Laboratory of Solid State Microstructures and 
Department of Physics, Nanjing University, Nanjing, P.R. China\\
3. Institute of Solid State Physics, Chinese Academy of Sciences,
Hefei, Anhui, China\\
4. Department of Physics, McGill University, Montreal, Quebec, 
Canada H3A 2T8}
\maketitle

\begin{abstract}
We report an exact solution for the noise spectrum of spin-current without 
charge-current in a spin field effect transistor (SFET). For the SFET with 
two leads, it is found that both auto- and cross-correlation functions 
are needed to characterize the noise of spin-current. A shot noise is also 
generated for the charge-current even though the net charge-current is 
identically zero. The noise spectra of spin-current and charge-current
depends on the coupling strength between the SFET and the leads, and they
may behave qualitatively differently. We also present results of noise
spectra in the adiabatic regime. 
\end{abstract}


Due to the particle nature of carriers, the fluctuation of charge-current 
gives rise to the notion of shot noise\cite{review1}. For a normal system, 
the current correlation between different probes (cross correlation) is 
negative for Fermions and positive for Bosons\cite{buttiker1,henny}. 
The correlation can be more complicated if superconducting leads are
present\cite{datta,torres,jehl,borlin,samu,wbg1,buttiker2}. Importantly,
the correlation properties provide further information, in addition to those
contained in conductance or charge-current, for mesoscopic 
conductors\cite{buttiker1}. So far all theoretical and experimental attention
have been devoted to correlations of charge-current.

In this paper, we theoretically investigate the correlation of 
{\it spin-current} in the absence of charge-current in a two-probe device. 
Indeed, to be able to generate and control spin-current is of great importance 
for spintronics\cite{ref1}, but the noise spectra of spin-current has never 
been studied before. A pure spin-current can behave qualitatively differently
from charge-current because spins can flip so that spin-current may not be
conserved. Therefore the cross correlation of currents may not be directly
related to the auto-correlation. On the other hand, the conservation of charge
forces the cross correlation and auto-correlation of charge-current noise 
spectra to differ by just a minus sign in a two-probe system. This difference 
necessarily indicates that the noise spectra of spin-current is more 
complicated. To be more specific, we will investigate the spin-current
correlation in a two-probe quantum spin field effect transistor 
(SFET)\cite{wbg2}. The detailed working principle of the SFET is summarized 
elsewhere\cite{wbg2} and we refer interested readers there: this device
delivers an instantaneous spin-current but without generating a net 
charge-current by the help of a rotating magnetic field. Roughly, if all the 
spin-up electrons in a conductor move to one direction while an equal 
number of spin-down electrons move to the opposite direction, 
the net charge-current vanishes because $I_e=e(I_{\uparrow}+I_{\downarrow})=0$;
and a finite spin-current results
because $I_s=\hbar/2(I_{\uparrow}-I_{\downarrow})\neq 0$. Here
$(I_{\uparrow},I_{\downarrow})$ are the electron current for spin-up and
spin-down electrons respectively. The SFET provides a mechanism for this
effect, and there are a number of other possibilities to produce $I_s \neq 0$ 
with $I_e=0$ which is the subject of several recently studies both
theoretically\cite{brataas,chamon,sun1} and experimentally\cite{marcus}.

For a two-lead device, correlations can be formed by quantities measured at 
the same lead---the auto-correlation, or by quantities measured at the two 
different leads---the cross-correlation. As discussed above, due to a lack
of spin-current conservation, we need both auto- and cross-correlation to
characterize the noise spectra of spin-current. It has been well
known\cite{prucell} that anti-bunching in a Fermionic system gives rise to 
negative definite cross-correlation for charge-current, and much of the
recent research has been devoted to situations where the sign of this
correlation\cite{review1} can be reversed.  For spin-current, the situation
is very different.  We found that when the coupling between SFET and lead 
is strong, the cross-correlation function of spin-current is positive definite.
In the intermediate or weak coupling regime, the cross-correlation of 
spin-current is positive when the system is far off a quantum resonance but 
it becomes negative near the resonance. In the adiabatic regime, the shot 
noise of a SFET involving only a single lead vanishes at resonance which 
corresponds to the quantization of delivered spin. As the coupling strength 
between the single lead 
and the SFET is varied, the shot noise is found to displays different number
of peaks depending on the coupling strength. Many of these features persist 
in the non-adiabatic regime. Finally, an oscillation between positive and 
negative shot noise is observed as a SFET parameter is tuned.  These
findings are qualitatively different and much richer than the behavior of
charge-current correlations.

We start by considering the SFET which generates spin-current without 
charge-current, described by the following Hamiltonian\cite{wbg2}
\begin{eqnarray}
&&H = \sum_{k,\sigma ,\alpha =L,R}\epsilon _{k}C_{k\alpha \sigma
}^{\dagger}C_{k\alpha \sigma }  
+\sum_{\sigma } [\epsilon +\sigma B_0 \cos\theta] d_{\sigma }^{\dagger}
d_{\sigma } \nonumber \\
&&+ \gamma [\exp(-i\omega t) d^\dagger_{\uparrow} d_{\downarrow} +h.c.]
+\sum_{k,\sigma,\alpha =L,R}[T_{k\alpha }C_{k\alpha \sigma
}^{\dagger}d_{\sigma }+h.c.] 
\label{hhh}
\end{eqnarray}
where the first term stands for the non-interacting electrons in 
lead $\alpha=L,R$ and $C_{k\alpha \sigma }^{\dagger}$ is the creation operator. 
Note that we have set the same chemical potential for the leads $L$ and 
$R$ because we are only interested in the quantum pumping effect\cite{pump} 
of the SFET. The second term describes the scattering region of the SFET
which is a quantum dot characterized by an energy level $\epsilon $ and 
spin $\sigma $. A gate voltage $v_g$ is applied to control the energy level.
The SFET generates spin-current\cite{wbg2} via a time-dependent magnetic field 
${\bf B}(t)=B_{0}[\sin \theta \cos \omega t ~ {\bf i}+
\sin \theta \sin \omega t ~ {\bf j}+\cos \theta {\bf k}]$. The last term of 
the Hamiltonian denotes tunneling between the leads and the dot with 
tunneling matrix elements $T_{k\alpha =L,R}$. Importantly,  we apply a
rotating magnetic field (the term proportional to $\gamma$) rather than an 
oscillating field: a counter-clockwise rotating field allows a spin-down 
electron to absorb a photon and flip to spin-up, and it does not allow a 
spin-up electron to emit a photon and flip to spin-down. As schematically
shown in the upper inset of Fig.(2), it is this symmetry breaking field 
that provides the driving force to deliver a pure DC spin-current to the
leads\cite{foot1}.

To analyze the noise spectrum of spin-current, we define a spin-dependent 
particle current operator ($\hbar=1$)
\begin{eqnarray}
\hat{J}_{\alpha ,\sigma } \equiv \sum_{k}\frac{d[C_{k\alpha \sigma
}^{\dagger}C_{k\alpha \sigma }]}{dt} 
=-i\sum_{k}[T_{k\alpha }C_{k\alpha \sigma }^{\dagger}d_{\sigma }-h.c.].
\label{J1}
\end{eqnarray}
Then the charge-current operator is
$\hat{I}_{\alpha q}=q\sum_{\sigma }\hat{J}_{\alpha,\sigma}$
An important quantity for our study is the correlation between
spin-dependent particle currents in lead $\alpha$ and $\beta$, 
\begin{eqnarray}
S_{\alpha \beta}^{\sigma \sigma'} 
= <[\hat{J}_{\alpha \sigma }(t_{1})-\bar{J}_{\alpha \sigma }] 
[\hat{J}_{\beta \sigma^{\prime }}(t_{2})-\bar{J}_{\beta \sigma^{\prime}}]>
\label{S1}
\end{eqnarray}
with $\bar{J}_{\alpha \sigma }\equiv <\hat{J}_{\alpha \sigma }>$ and
$\sigma,\sigma^{\prime }$ denoting spin indices. Here $<\cdots >$
denotes both statistical average and quantum average on the nonequilibrium 
state. The noise spectra of both charge-current and spin-current can be 
obtained from the correlation $S_{\alpha \beta}^{\sigma \sigma'}$.

We calculate $S_{\alpha \beta}^{\sigma \sigma'}$ using standard Keldysh
nonequilibrium Green's function (NEGF) formalism\cite{wbg2}. Briefly, 
we substitute Eq.(\ref{J1}) into Eq.(\ref{S1}), define NEGF $G^<$ and $G^>$
with properly contour ordered 
operators, and apply the theorem of analytic continuation\cite{wbg1} so that
the contour Green's functions are extended to the real time axis. This
standard and widely used NEGF technique allows us to obtain the exact
expression for the zero-frequency spin-dependent correlation\cite{foot5}.

We now examine noise spectrum in the low temperature limit $kT<<\hbar\omega$, 
{\it i.e.} the shot noise. First, setting $\alpha=$L, $\beta=$R in
$S_{\alpha \beta}^{\sigma \sigma'}$, we investigate shot 
noise of the charge-current (cross correlation).  It has been shwon in
Ref.\onlinecite{wbg2} that the net charge-current of the SFET is identically
zero\cite{foot1}, however here we find that shot noise of it is nonzero. 
The shot noise of the charge-current is
\begin{eqnarray}
&&S_{ele}=<\Delta I_L \Delta I_R>=q^2\sum_{\sigma \sigma'} 
S_{LR}^{\sigma \sigma'} \nonumber \\
&&=-q^2 \Gamma _{L}\Gamma _{R}\int \frac{dE}{2\pi }[\mid G_{\uparrow
\downarrow }^{r}\mid ^{2}+\mid G_{\downarrow \uparrow }^{r}\mid
^{2}]f_{\downarrow }(1-f_{\uparrow })
\end{eqnarray}
where $G^r_{\sigma \bar{\sigma}}=\gamma/[ (E-\epsilon+i\Gamma/2) (E-\epsilon
+\sigma \omega+i\Gamma/2)-\gamma^2]$ is obtained from Ref.\onlinecite{wbg2}. 
Here $\Gamma_\alpha$ is the linewidth function, $f_{\uparrow}=f_{\uparrow}
(E)$,  and $f_{\downarrow} =f_{\downarrow}(E-\omega)$. We conclude that 
$S_{ele}\neq 0$ although the instantaneous 
charge-current is zero\cite{wbg2}. The excess noise has the following 
lineshape in the adiabatic limit ({\it i.e.} $\omega \rightarrow 0$),
\begin{equation}
S_{ele}=-\frac{q^2 \omega}{2\pi}\frac{2\Gamma_L \Gamma_R \gamma^2}
{(\epsilon^2+\Gamma^2/4-\gamma^2)^2+\Gamma^2 \gamma^2}
\label{s1}
\end{equation}
here $\epsilon$ is the resonant level of the quantum dot and we have 
set $\theta=\pi/2$ and $E_F=0$. As expected, the excess noise (cross
correlation) in charge-current is always negative. A single peak is found at 
$\epsilon=0$ for $\gamma \leq \Gamma/2$ while a double peak structure occurs
when $\gamma > \Gamma/2$. This can be understood as follows. When the spin 
flip magnetic field $\gamma$ is turned on, the spin degeneracy of resonant 
state is broken. Two resonant states are found at 
$\epsilon\pm \gamma -i\Gamma/2$ with level spacing $2\gamma$ and
half width $\Gamma/2$. When $\gamma > \Gamma/2$, the two resonant states
do not overlap and we obtain two peaks. However, if $\gamma \leq \Gamma/2$,
only one peak is showed up due to the overlapping of resonant states. 
Finally, the auto-correlation is obtained by setting $\alpha=\beta=$L in
$S_{\alpha \beta}^{\sigma \sigma'}$, and it is straightforward to show 
$<\Delta I_L \Delta I_L> =
-<\Delta I_L \Delta I_R>$. This is the expected result for charge-current
which is a conserved quantity, {\it i.e.} $I_L+I_R=0$. 

Next, we calculate the shot noise for spin-current. 
The cross correlation spin-current shot noise is found to be:
\begin{eqnarray}
&&S_{spin,1}=<(\Delta J_{L\uparrow }-\Delta J_{L\downarrow })
(\Delta J_{R\uparrow }-\Delta J_{R\downarrow })>= \nonumber \\
&&(S_{LR}^{11}+S_{LR}^{22}-S_{LR}^{12}-S_{LR}^{21})/4 = 
\int \frac{dE}{8\pi }f_{\downarrow }(1-f_{\uparrow }) \nonumber \\
&&\Gamma_L \Gamma_R [ |G^r_{\uparrow \downarrow}|^2 
+|G^r_{\downarrow \uparrow}|^2 -2 \Gamma^2 ( |G^r_{\uparrow 
\downarrow}|^4 +|G^r_{\downarrow \uparrow}|^4 )]
\label{SS1}
\end{eqnarray}
and the auto-correlation is found to be:
\begin{eqnarray}
&&S_{spin,2}=<(\Delta J_{L\uparrow }-\Delta J_{L\downarrow })
(\Delta J_{L\uparrow }-\Delta J_{L\downarrow })> \nonumber \\
&&=(S_{LL}^{11}+S_{LL}^{22}-S_{LL}^{12}-S_{LL}^{21})/4 = 
\int \frac{dE}{8\pi }f_{\downarrow }(1-f_{\uparrow }) \nonumber \\
&&\left\{2[-\Gamma _{L}^{2}\Gamma ^{2} (|G_{\uparrow \downarrow }^{r}|^4
+|G_{\downarrow \uparrow }^{r}|^4) 
+\Gamma _{L}\Gamma (|G_{\uparrow \downarrow }^{r}|^2 \right.
\nonumber \\
&&\left.+|G_{\downarrow \uparrow }^{r}|^2) ]-\Gamma _{L}\Gamma _{R}
(|G_{\uparrow \downarrow }^{r}|^2+|G_{\downarrow \uparrow }^{r}|^2)\right\}
\label{SS2}
\end{eqnarray}
In contrast to the shot noise of charge-current, we need both the cross- 
and the auto-correlation to characterize the shot noise for spin-current for
the two-lead SFET. The reason is because a spin-current is not conserved due
to the spin flip mechanism introduced by the rotating magnetic field.
In the adiabatic limit, the cross correlation reduces to 
\begin{eqnarray}
&&S_{spin,1}=\frac{\omega}{4\pi}\frac{\Gamma_L \Gamma_R \gamma^2
(\epsilon^2+\Gamma^2/4-\gamma^2+\Gamma \gamma)} 
{[(\epsilon^2+\Gamma^2/4-\gamma^2)^2+\Gamma^2 \gamma^2]^2} \nonumber \\
&&\times (\epsilon^2+\Gamma^2/4-\gamma^2-\Gamma \gamma)
\label{SS1a}
\end{eqnarray}
Hence, cross correlation $S_{spin,1}$ can be either positive or negative 
depending on a number of parameters: the gate voltage which controls the 
energy level position, the linewidth function $\Gamma$, and the external 
magnetic field strength $\gamma$ (see Fig.1). Far away from resonance, 
the cross correlation is always positive when $x>\sqrt{2}$ 
(with $x\equiv |\epsilon|/\gamma$) regardless of the coupling strength between 
quantum dot and the leads. On the other hand, for $x<\sqrt{2}$, the cross 
correlation is positive definite when the coupling of the lead to the 
quantum dot is strong such that $\Gamma >2(1+\sqrt{2-x}) \gamma$. 
When the coupling to the lead is in the intermediate range 
$2|\sqrt{2-x}-1| \gamma <\Gamma <2(1+\sqrt{2-x})\gamma$,
the cross correlation turns to negative. In the weakly coupled regime 
$\Gamma < 2|\sqrt{2-x}-1|\gamma$, the cross correlation becomes 
positive again. The cross correlation is zero when $\Gamma = 2(\sqrt{2-x} 
+1)\gamma$ or $\Gamma = 2|\sqrt{2-x}-1| \gamma$ with $x \leq 2$.  
Interestingly, as one varys the ratio $\Gamma/(2\gamma)$, different
lineshapes are found for the cross correlation: (1) in the strong
coupling regime (compared with 2$\gamma$): $\Gamma/2>(2+\sqrt{3})\gamma$,
the cross correlation is positive definite with a broad peak at 
$\epsilon=0$ (see inset of Fig.1); (2) As one decreases the coupling
strength $\Gamma$ such that $\gamma <\Gamma/2<(2+\sqrt{3}) \gamma$, the
positive peak at $\epsilon=0$ becomes a local minimum and a double peak 
structure appears. This can also be seen in the shot noise for the 
charge-current (auto correlation)\cite{ref}. Due to 
the overlap of the two resonant states, two peaks are found (solid line in
Fig.1); (3) As one further decreases $\Gamma$, a third peak emerges at
$\epsilon=0$ when $\gamma >\Gamma/2>(2-\sqrt{3}) \gamma$ (so that two 
resonant states do not overlap) (see dotted line in Fig.1); (4) finally, 
in the weak coupling regime ($\Gamma/2<(2-\sqrt{3})\gamma$), the third 
peak splits and a four-peak structure appears in the shot noise (dashed 
line in Fig.1). Because of the spin flip mechanism, both spin-up and 
spin-down electrons are contributing to the spin current. The cross correlation 
between spin up electrons (or spin down) is negative definite while 
the cross correlation between spin-up electron and spin-down electron
is positive definite. The competition between these two contributions 
gives rise to either a positive or a negative cross correlation. 
Such a complicated behavior is qualitatively different from the correlation
in charge-current. 

In the adiabatic regime, the auto-correlation is obtained from 
Eq.(\ref{SS2}),
\begin{equation}
S_{spin,2}=\frac{\omega\Gamma_L \gamma^2 [(\Gamma_L
+\Gamma) (\epsilon^2+\Gamma^2/4-\gamma^2)^2+\Gamma_R \Gamma^2
\gamma^2]} 
{4\pi[(\epsilon^2+\Gamma^2/4-\gamma^2)^2+\Gamma^2 \gamma^2]^2} 
\label{s3}
\end{equation}
which is positive definite. When the cross correlation is zero, the auto
correlation is $\omega \Gamma_L /(8\pi\Gamma)$. 

The SFET can operate with a single lead\cite{wbg2}.  In this case, the shot 
noise is found to be 
\begin{equation}
S_{spin}=\frac{\omega\Gamma^2 \gamma^2 (\epsilon^2+\Gamma^2/4-\gamma^2)^2} 
{2\pi[(\epsilon^2+\Gamma^2/4-\gamma^2)^2+\Gamma^2 \gamma^2]^2} 
\label{s4}
\end{equation}
We observe that this shot noise is positive definite and it displays a 
single to four-peak structure in the lineshape as $\Gamma$ is changed.  This
is similar to the behavior of $S_{spin,1}$. When $\gamma>\Gamma/2$, the shot 
noise is zero at $\epsilon= \pm \sqrt{\gamma^2-\Gamma^2/4}$.
For a single lead SFET, the spin current was found before\cite{wbg2}, 
\begin{equation}
I_s=-\frac{\omega}{2\pi}\frac{\Gamma^2 \gamma^2}
{(\epsilon^2+\Gamma^2/4-\gamma^2)^2+\Gamma^2 \gamma^2}
\end{equation}
When the shot noise is identically zero, $I_s=\omega/(2\pi)$. This means 
that the SFET pumps out two spin quanta ($\hbar=1$) in a period 
$\tau=2\pi/\omega$. This gives an example of ``optimal'' spin pump.  The  
analogous optimal charge pump\cite{avron} was discussed in literature 
which is actually rather difficult to achieve\cite{levinson,wbg3}. 

Our exact solutions Eqs.(\ref{SS1}) and (\ref{SS2})
also allow the investigation of
shot noise of spin-current in the non-adiabatic regime ($\omega\neq 0$). 
We fix field strength $\gamma=0.02$ in the following calculation.
When frequency $\omega\neq 0$, the noise becomes asymmetric with 
respect to the gate voltage. 
For $\Gamma=0.017$, the influence of frequency is shown in the left inset 
of Fig.2. We observe that when $\omega=0.01$ (solid line), the depths of 
the two minima decrease. As frequency changes to $\omega=0.02$ (dotted line), 
the noise is still negative for a wide range of gate voltages and 
is positive for some negative gate voltages. Finally, when $\omega=0.03$, 
the noise becomes positive definite. The noise as a function of frequency
is depicted in Fig.2. The oscillatory behavior between positive and negative 
noise of the spin-current is observed, which is due to photon assisted 
process. 

In summary, we have, for the first time, presented an exact solution of
the shot noise for spin-current generated by the SFET which is essentially
a quantum spin pump at arbitrary pumping frequency. For the device discussed
here, the instantaneous charge current is zero but its shot noise is
finite. Due to the spin flip mechanism in the SFET, the spin-current is not 
conserved. As a result, both auto-correlation and cross correlation must 
be used to characterize the shot noise of spin-current for our two-lead 
system. Away from a quantum resonance, we found that the cross correlation 
becomes positive due to the occurance of carriers with different spin. 
In the adiabatic regime, the auto-correlation of a quantum dot connected to
a single lead is calculated and found to be zero when the pumped spin 
is quantized. As the coupling strength between the single lead and the 
quantum dot is varied, we found that the shot noise displays different 
lineshapes in the different coupling regimes. 

\section*{Acknowledgments}
We gratefully acknowledge support by a RGC grant from the SAR Government of 
Hong Kong under grant number HKU 7113/02P
and from NSERC of Canada and FCAR of Quebec (H.G).

\bigskip
\bigskip
\bigskip
\noindent{$^{a)}$ Electronic mail: jianwang@hkusub.hku.hk}

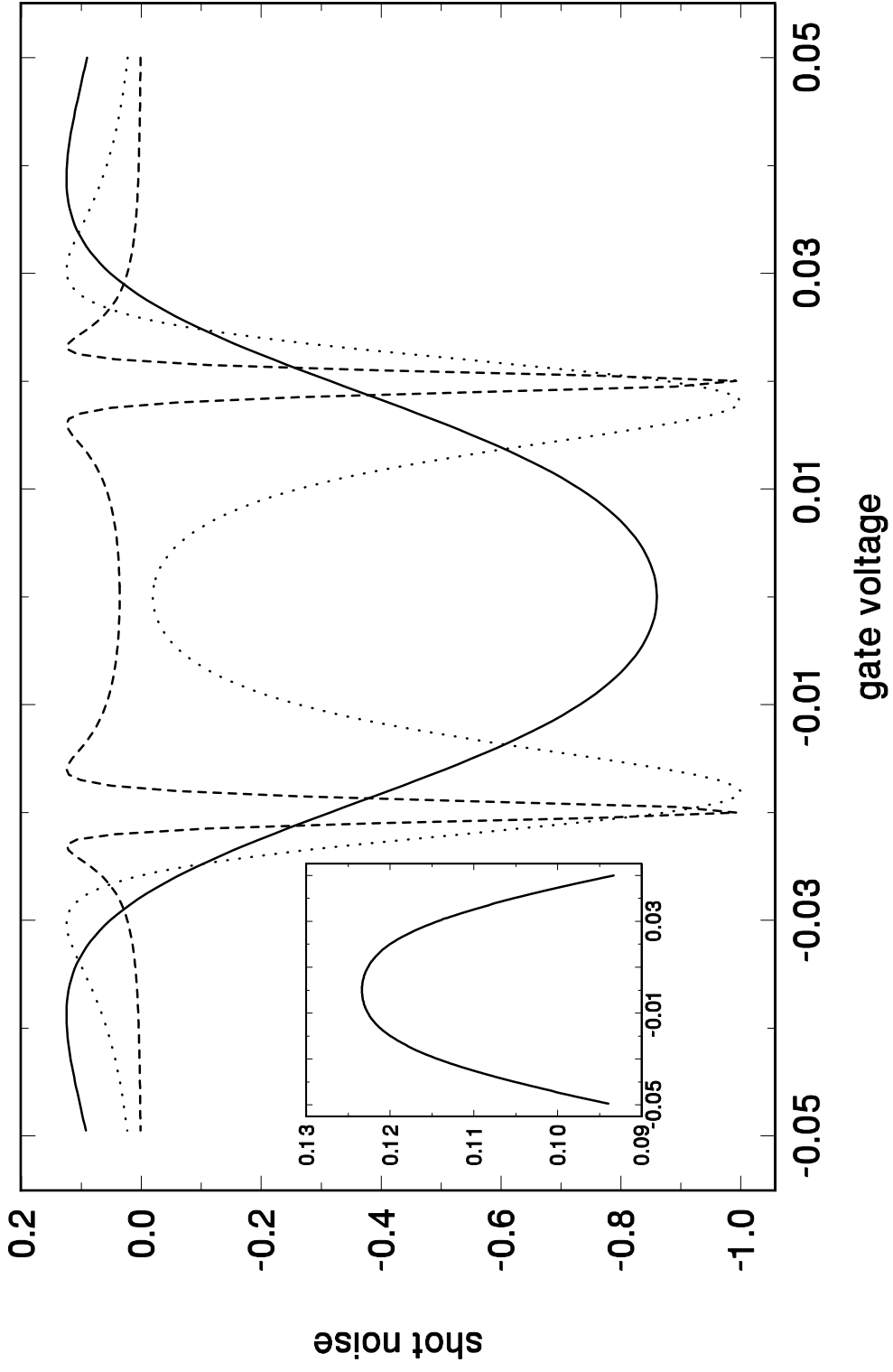
\begin{figure}
\caption{
The shot noise for spin current $S_{spin,1}$ versus the gate voltage 
$v_g$ for different coupling strength: (1). $\Gamma=0.05$ (solid line); 
(2). $\Gamma=0.017$ (dotted line); (3). $\Gamma=0.004$ (dashed line). 
Inset: $S_{spin,1}$ vs $v_g$ when $\Gamma=0.16$. Here $\gamma=0.02$,
$\Gamma_l=\Gamma_R$, and
the shot noise is plotted in the unit of $\omega/(16\pi)$.
}
\end{figure}

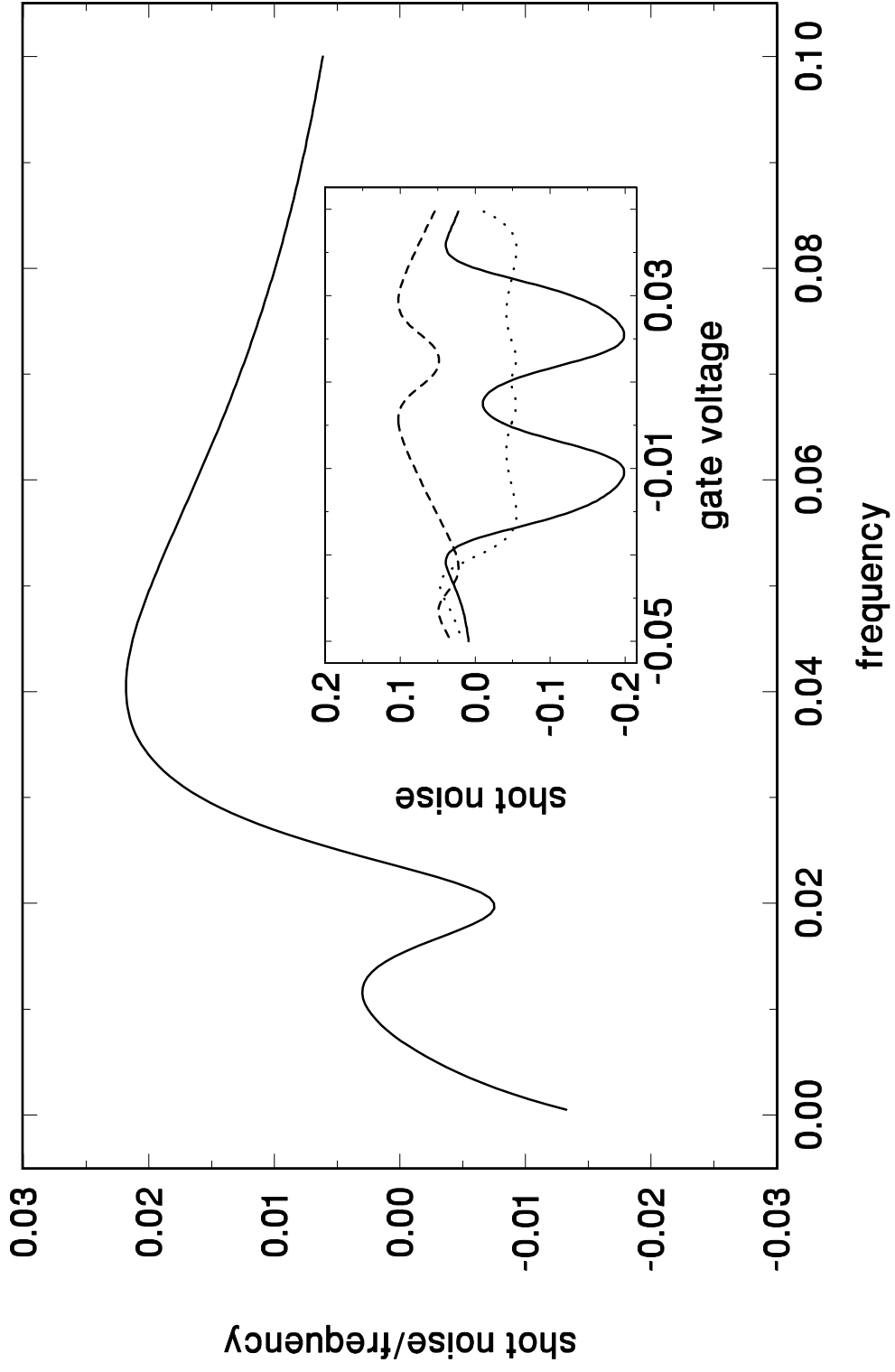
\begin{figure}
\caption{
$S_{spin,1}/\omega$ versus frequency for $v_g=0.007$ and $\Gamma=0.016$. 
The unit is $1/(4\pi)$.  Inset: $S_{spin,1}$ versus gate voltage
at different frequency when $\Gamma=0.017$: $\omega=0.01$ (solid line); 
$\omega=0.02$ (dotted line); $\omega=0.03$ (dashed line). 
The unit is $\omega/(4\pi)$ with $\omega=0.005$.
}
\end{figure}


\begin{thebibliography}{00} 

\bibitem{review1}
Ya. M. Blanter and M. Buttiker, Phys. Rept. {\bf 336}, 1 (2000);
M. Buttiker, cond-mat/0209031.

\bibitem{buttiker1}
M. Buttiker, Phys. Rev. B {\bf 46}, 12485 (1992).

\bibitem{brown}
R. Hanbury Brown and R.Q. Twiss, Nature (London), {\bf 177}, 27 (1956).

\bibitem{henny}
M. Henny et al., Science, {\bf 284}, 296 (1999). 

\bibitem{oliver}
W.D. Oliver et al., Science, {\bf 284}, 299 (1999).

\bibitem{datta}
M.P. Anantram and S. Datta, Phys. Rev. B {\bf 53}, 16390 (1996).

\bibitem{torres}
J. Torres and Th. Martin, Eur. Phys. J. B {\bf 12}, 319 (1999).

\bibitem{jehl}
X. Jehl et al, Nature {\bf 405}, 50
(2000); A.A. Kozhevnikov et al, Phys. Rev. Lett. {\bf 84}, 3398 (2000).

\bibitem{borlin}
%J. Borlin, W. Belzig, and C. Bruder, Phys. Rev. Lett. {\bf 88}, 197001
J. Borlin et al, Phys. Rev. Lett. {\bf 88}, 197001 (2002). 

\bibitem{samu}
P. Samuelsson and M. Buttiker, Phys. Rev. Lett. {\bf 89}, 046601
(2002). 

\bibitem{wbg1}
B.G. Wang and J. Wang, Phys. Rev. B {\bf 67}, 014509 (2003).

\bibitem{buttiker2}
A.M. Martin and M. Buttiker, Phys. Rev. Lett. {\bf 84}, 3386 (2000).

\bibitem{ref1}
S.A. Wolf {\it et.al.}
%D.D. Awschalom, R.A. Buhrman, J.M. Daughton, S.V. Molnar,
%M.L. Roukes, A.Y. Chtchelkanova, D.M. Treger, 
Science {\bf 294}, 1488 (2001); G.A. Prinz, Science {\bf 282}, 1660 (1998).

\bibitem{wbg2}
B.G. Wang, J. Wang, and H. Guo, Phys. Rev. B {\bf 67}, 092408 (2003).

\bibitem{brataas}
%A. Brataas, Y. Tserkovnyak, G.E.W. Bauer, and B.I. Halperin,
%cond-mat/0205028.
A. Brataas {\it et.al.}, Phys. Rev. B {\bf 66}, 060404 (2002).

\bibitem{chamon}
P. Sharma and C. Chamon, Phys. Rev. Lett. {\bf 87}, 096401(2001);
%E.R. Mucciolo, C. Chamon and C.M. Marcus,
E.R. Mucciolo et al, {\it ibid}. {\bf 89}, 146802(2002).

\bibitem{sun1}
Q.-f. Sun, H. Guo, and J. Wang, cond-mat/0212293 (2002).

\bibitem{marcus}
Susan K. Watson, R.M. Potok, C.M. Marcus and V. Umansky, cond-mat/0302492
(2003).

\bibitem{prucell}
E.M. Prucell, Nature, {\bf 178}, 1449 (1956).

\bibitem{pump}
P.W. Brouwer, Phys. Rev. B {\bf 58}, R10135 (1998); F. Zhou et al, 
Phys. Rev. Lett. {\bf 82}, 608 (1999); T.A. Shutenko et al, Phys. Rev. B 
{\bf 61}, 10366 (2000); Y.D. Wei et al, {\it ibid} {\bf 62}, 9947 (2000); 
M. Moskalets and M. Buttiker, {\it ibid} {\bf 64}, 201305 (2001); B.G.
Wang et al, {\it ibid} {\bf 65}, 073306 (2002).

\bibitem{foot1}
Briefly, the SFET works as follows. The z-component of field ${\bf B}(t)$ 
splits the level $\epsilon$ into $\epsilon_{\downarrow}<\epsilon_{\uparrow}$.
The chemical potential $\mu$ of the leads is adjusted so that 
$\epsilon_{\downarrow}<\mu<\epsilon_{\uparrow}$ as shown in the upper inset
of Fig.(2). A spin-down electron can tunnel into $\epsilon_{\downarrow}$
from the left lead, it absorbs a photon due to the rotating field and
flip its spin to occupy $\epsilon_{\uparrow}$. Because
$\epsilon_{\uparrow}> \mu$, the now spin-up electron tunnels out to the leads
easily.  The same process occurs to spin-down electrons coming from the
right lead.  Therefore, spin-down electrons flow toward the quantum dot
while spin-up electrons electrons flow away from the dot, giving rise to a
zero net charge-current and a finite spin-current.  See
Ref.\onlinecite{wbg2} for more details.

\bibitem{foot5}
The detailed derivation will be published elsewhere.
 
\bibitem{ref}
Y.D. Wei, B.G. Wang, J. Wang, and H. Guo 
%Y.D. Wei et al, 
Phys. Rev. B {\bf 60}, 16900 (1999). 

\bibitem{avron}
%J.E. Avron, A. Elgart, G.M. Graf, and L. Sadun, Phys. Rev. Lett. 
%{\bf 87}, 236601 (2001).
J.E. Avron et al, Phys. Rev. Lett. {\bf 87}, 236601 (2001).

\bibitem{levinson}
%Y. Levinson, O. Entin-Wohlman, and P. Wolfle, Physica A {\bf 302},
%335 (2001). 
Y. Levinson et al, Physica A {\bf 302}, 335 (2001). 

\bibitem{wbg3}
B.G. Wang and J. Wang, Phys. Rev. B {\bf 66}, 125310 (2002). 

\end{thebibliography}
\end{document}